\documentclass{ws-jtca}

\usepackage[super]{cite}
\usepackage[colorlinks = true,
            linkcolor = blue,
            urlcolor  = blue,
            citecolor = blue,
            anchorcolor = blue]{hyperref}
\newcommand{\twovec}[2]{\begin{pmatrix} #1 \\ #2 \end{pmatrix}}
\newcommand{\pd}[2]{\frac{\partial #1}{\partial #2}}
\newcommand{\pds}[2]{\frac{\partial^2 #1}{\partial #2^2}}
\newcommand{\eq}[1]{\begin{equation} #1 \end{equation}}

\newcommand{\ux}{u_x}
\newcommand{\sigmaxx}{\sigma_{xx}}
\newcommand{\sigmaxz}{\sigma_{xz}}

\makeatletter
\def\blfootnote{\xdef\@thefnmark{}\@footnotetext}
\makeatother

\begin{document}

\markboth{M. D. Collins and A. Ramamurti}{Approaches for handling sloping fluid-solid interfaces with the parabolic equation method}

%
\catchline{}{}{}{}{}
%

\title{Approaches for handling sloping fluid-solid interfaces with the parabolic equation method}
 
\author{Michael D. Collins \footnote{\email{michael.collins@nrl.navy.mil} \vspace{-3ex}}}
\address{Code 7160, Acoustics Division,\\ U.S. Naval Research Laboratory,\\ Washington, D.C. 20375}

\author{Adith Ramamurti\footnote{\email{adith.ramamurti@nrl.navy.mil}}}
\address{Code 7165, Acoustics Division,\\ U.S. Naval Research Laboratory,\\ Washington, D.C. 20375}

\maketitle


\begin{abstract}
Several methods for handling sloping fluid-solid interfaces with the elastic parabolic equation are tested. 
A single-scattering approach that is modified for the fluid-solid case is accurate for some problems but breaks down when the contrast across the interface is sufficiently large and when there is a Scholte wave. 
An approximate condition for conserving energy breaks down when a Scholte wave propagates along a sloping interface but otherwise performs well for a large class of problems involving gradual slopes, a wide range of sediment parameters, and ice cover.
An approach based on treating part of the fluid layer as a solid with low shear speed handles Scholte waves and a wide range of sediment parameters accurately, but this approach needs further development. 
The variable rotated parabolic equation is not effective for problems involving frequent or continuous changes in slope, but it provides a high level of accuracy for most of the test cases, which have regions of constant slope.
Approaches based on a coordinate mapping and on using a film of solid material with low shear speed on the rises of the stair steps that approximate a sloping interface are also tested and found to produce accurate results for some cases. 
\end{abstract}

\section{Introduction}

The parabolic equation method\cite{jensen1994,collins2019a} provides an attractive combination of accuracy and efficiency for many wave propagation problems in which the parameters of the medium have strong variations in depth and gradual range dependence (variations in the horizontal directions). 
For problems in ocean acoustics, Arctic acoustics, and seismology, range dependence may including sloping interfaces, sloping boundaries, and continuous variations within the interiors of layers. 
Existing approaches for handling range dependence in elastic parabolic equation solutions are based on energy conservation, single scattering, and coordinate transformations.\cite{collins2019a} 
Effective approaches have been developed for handling sloping solid-solid interfaces and sloping solid boundaries.\cite{woolfe2016a} 
Some progress has been made in the treatment of sloping fluid-solid interfaces,\cite{woolfe2016b,fialkowski2018,collins2019b} but there is a need for improvement in this area. 
In this paper, several approaches for handling sloping fluid-solid interfaces are discussed and tested, including an approach based on modeling part of a fluid layer as a solid material with low shear wave speed. 

\section{Approaches for Handling Range Dependence}

A derivation of the elastic parabolic equation is outlined here in Cartesian coordinates, where the range $x$ is the horizontal distance from a line source and $z$  is the depth below the top boundary; see Ref.~\citen{collins2019a} for further details. 
Sloping boundaries and interfaces and other types of horizontal variations in the properties of the medium may be approximated in terms of a series of range-independent regions in which the properties vary only with depth. 
The parabolic equation method is based on the assumption that outgoing energy (which propagates away from the source in the positive $x$ direction) dominates back scattered energy (which propagates toward the source in the negative $x$ direction). 
In each range-independent region, the elastic wave equation is in the form,
\eq{
\left( L \pds{}{x} + M \right) \twovec{\ux}{w} = 0\,,
\label{eq:elasticWE}
}
where $u_x$ is the horizontal derivative of the horizontal displacement, $w$  is the vertical displacement, and the entries of the $2\times2$ matrices  $L$ and $M$  are depth operators (second-order differential operators plus interface and boundary conditions). 
Factoring the operator in Eq.~(\ref{eq:elasticWE}) and assuming that outgoing energy dominates, we obtain the parabolic wave equation,
\begin{equation}
\pd{}{x} \twovec{u_x}{w} = i(L^{-1} M)^{1/2}  \twovec{\ux}{w} \,.
\label{eq:elasticPE}
\end{equation}
Numerical solutions of Eq.~(\ref{eq:elasticPE}) may be obtained by constructing an initial condition using the self starter, approximating the operator square root with a rational function, and applying numerical approximations to march the solution through each range-independent region.\cite{collins2019a} 

At a vertical interface between two range-independent regions, the exact solution satisfies continuity of $u$, $w$, the normal stress $\sigmaxx$, and the tangential stress $\sigmaxz$. 
Since the number of range derivatives is reduced in going from Eq.~(\ref{eq:elasticWE}) to Eq.~(\ref{eq:elasticPE}), it is not possible for a parabolic equation solution to satisfy all four conditions. 
Various approaches have been developed for handling vertical interfaces, including the single-scattering approximation,\cite{collins2012}
\eq{
\twovec{\ux}{w}_t = \frac{i}{2} (L_B^{-1}M_B)^{1/2} S_B^{-1}\twovec{u}{-\sigmaxz}_i +\frac{1}{2}R_B^{-1}\twovec{\sigmaxx}{w}_i\,,
\label{eq:ss1}
}
\eq{
\twovec{\sigmaxx}{w}_i = R_A\twovec{\ux}{w}_i\,,
\label{eq:ss2}
}
\eq{
\twovec{u}{-\sigmaxz}_i = -iS_A(L_A^{-1}M_A)^{-1/2}\twovec{\ux}{w}_i\,,
\label{eq:ss3}
}
where $R$ and $S$ are depth operators, the subscripts $i$ and $t$ denote the incident and transmitted fields, and the subscripts $A$ and $B$ denote the regions on the incident and transmitted sides of the vertical interface. 
This approach provides accurate solutions for many problems involving sloping fluid-fluid interfaces, sloping solid-solid interfaces, and sloping solid boundaries.\cite{woolfe2016a} 

Various approaches have been developed for handling sloping fluid-solid interfaces, but none of them is as effective as Eq.~(\ref{eq:ss1}) is for the other cases. 
Since the tangential displacement is not continuous across a fluid-solid interface, Eq.~(\ref{eq:ss1}) is not appropriate for the fluid-solid case, but the following modification is effective for some problems:
\eq{
\twovec{\ux}{w}_t = \frac{i}{2}(L_B^{-1} M_B)^{1/2} S_B^{-1}\twovec{u}{-\sigmaxz}_i + \frac{1}{2}Q_B^{-1}\twovec{\sigmaxx}{-\partial \sigmaxz / \partial x}_i\,,
\label{eq:modss}
}
\eq{
\twovec{\sigmaxx}{-\partial \sigmaxz / \partial x}_i = Q_A\twovec{\ux}{w}_i\,,
}
where $Q$ is a depth operator.\cite{woolfe2016b} 
Conditions for conserving energy flux have proven to be effective for the acoustic case. 
In its most general form for the acoustic case, the energy-conservation condition identifies the incident field as an array of point sources with the correct energy-flux densities.
When generalized to the elastic case, the energy-conservation condition identifies the incident field as arrays of compressional and shear point sources with the correct energy-flux densities.\cite{collins1999}
This approach has not yet been fully implemented (or demonstrated to be stable and accurate), but the following approximate implementation provides accurate solutions for many problems involving sloping fluid-solid interfaces:\cite{collins2019b}
\eq{
\rho_B^{1/2}S_B(L_B^{-1}M_B)^{-1/2}\twovec{\ux}{w}_t = \rho_A^{1/2}S_A(L_A^{-1}M_A)^{-1/2}\twovec{\ux}{w}_i\,. 
\label{eq:appec}
}
The first row of this condition is related to the condition for conservation of energy for the compressional wave; the second row correspond to vanishing tangential stress at the fluid-solid interface. 
This condition provides perhaps the most attractive combination of accuracy and stability among the currently existing approaches for handling sloping fluid-solid interfaces. If the density factors are removed, Eq.~(\ref{eq:appec}) reduces to a condition for conservation of $u$ and $\sigmaxz$. 

Sloping fluid-solid interfaces may also be handled by mapping\cite{collins2000} or rotating\cite{outing2006} coordinates. 
The mapping approach is based on a change of variables in which the environment is rigidly translated vertically at each range so that a sloping fluid-solid interface becomes a horizontal interface. 
Under this mapping, additional terms are introduced in the wave equation and the surface becomes a sloping boundary. 
For small slopes, the additional terms may be neglected or approximately taken into account with a correction factor for the phase.\cite{collins2000} 
The tradeoff of a sloping fluid-solid interface for a sloping boundary is a bargain since the latter is easier to handle. 
The variable rotated parabolic equation is based on coordinate systems that are rotated to be aligned with the local slope of the interface. 
When there is a change in slope, the solution is advanced slightly beyond the change in slope, and an initial condition is constructed for the rotated coordinate system in the next region by interpolation and extrapolation.\cite{outing2006} 
For the case of a solid medium, it is necessary to rotate the dependent variables at each change in slope. 

For some problems, accurate solutions may be obtained by introducing an artificial layer of solid material with a low shear speed. 
For some downslope problems, accurate solutions have been obtained by introducing a thin film on the rises of the stair steps that approximate a sloping fluid-solid interface and solving two scattering problems.\cite{woolfe2016b} 
At least three grid points are required on the rise when this approach is implemented using a non-centered four-point difference formula for a second derivative that appears in one of the conditions for a horizontal fluid-solid interface.\cite{collins2015} 
An alternate approach that does not require multiple grid points on the rises is to approximate part of the fluid layer with a low shear speed solid and use Eq.~(\ref{eq:ss1}) to handle what becomes a sloping solid-solid interface. 
Going from the elastic wave equation to the acoustic wave equation by allowing the shear speed to approach zero is a singular limit. 
At a horizontal interface between a solid and a low shear speed solid that is used to model a fluid, the tangential displacement is continuous across the interface for any finite value of the shear speed, but this quantity is not continuous across a fluid-solid interface. 
One would therefore expect the solution to vary rapidly in a boundary layer near the interface.
One approach for implementing this solution efficiently would be to use variable grid spacing, with relatively fine spacing in the boundary layer to account for rapid variations.
Another possible approach would be to use slip conditions at the interface so that the tangential displacement is not conserved; this approach might eliminate the rapid variations and the requirement for a fine grid. 

\section{Test Cases}

We consider several examples for testing the approaches for handling sloping fluid-solid interfaces with the seismo-acoustic parabolic equation. 
For each case, the sloping interface has the same 2.86 degree slope that was used for a series of benchmark problems for the acoustic parabolic equation.\cite{jensen1990} 
The parabolic equation method can handle larger slopes, but accuracy will degrade for a sufficently large slope. 
The parabolic equation solutions are obtained using rotated rational approximations\cite{millinazzo1997} for the operator square root. 
The usual definition for transmission loss is used in the water column. 
The transmission loss in solid layers is defined in terms of $\sigma_{zz}$. 
The examples are intended to serve as convenient test cases for the seismo-acoustic parabolic equation but are not intended to be realistic. 

A finite-element model\cite{comsol} is used to generate reference solutions. 
Such models typically require substantially greater computation times than parabolic equation solutions, but their efficiency may be improved by applying an iteration scheme rather than a direct solver of a large system of equations. 
The primary objective here is to test the accuracy of approaches for handling sloping fluid-solid interfaces, but a baseline efficiency comparison was performed. 
For a range-independent version of one of the problems considered here, the run time of the finite-element model is 91.5 times greater than the run time of the parabolic equation model. 
It takes longer to solve range-dependent problems with the parabolic equation method, but the increased run time is typically less than a factor of two if the matrices involved in the numerical solution are updated efficiently as the parameters vary with range. 

The parabolic equation is solved with a marching approach, in which a small fraction of the grid must be stored in memory at any time. 
Since a finite-element solution is obtained over the entire grid simultaneously, the difficulty of this approach rapidly increases with the size of the grid. 
Although finite-element models may in principle be used to produce solutions of arbitrary accuracy, it can be difficult in practice to confirm that an accurate solution has been obtained. 
In designing the test cases in Ref.~\citen{woolfe2016b}, for example, the size of the range-depth domain was chosen to be near the limit of what could be handled with the finite-element model on the computer that was used in that study; one of the motives for this choice was to select problems that are sufficiently large to be useful and interesting test cases. 
Convergence tests were conducted, but there were uncertainties in some of the solutions. 
During the tests presented here, which were conducted on a computer with greater capacity, it was found that there are noticeable errors in some of the finite-element solutions appearing in Ref.~\citen{woolfe2016b}. 

Each of the test cases involves a 25~Hz point source in cylindrical geometry. 
All layers are homogeneous. 
The sound speed is 1500~m/s in the water column. 
We consider the six test cases that appear in Figs.~3--8 of Ref.~\citen{woolfe2016b} plus an additional test case that involves an interface wave. 
For the test cases of Ref.~\citen{woolfe2016b}, the source is 380~m below the surface, and the bathymetry is 400~m for $r<3$~km, linearly varying by 200~m over 3~km~$<r<7$~km  (downslope and upslope cases), and constant for $r>7$~km. 
We consider three sediment types for which the compressional and shear attenuations are both  0.2~dB$/\lambda$. 
For the low-contrast case, the compressional speed is 1700~m/s, the shear speed is 800 m/s, and the density is  1.2~g/cm$^3$. 
For the medium-contrast case, the compressional speed is 2400~m/s, the shear speed is~1200 m/s, and the density is  1.5~g/cm$^3$. 
For the high-contrast case, the compressional speed is 3400~m/s, the shear speed is 1700~m/s, and the density is 2.5~g/cm$^3$. 
For the problem involving an interface wave, the bathymetry decreases from 500~m to 300~m over the same range interval as in the other cases, the source is placed 490~m below the surface to excite a Scholte wave along the interface, and the low-contrast sediment is used with one modification: the compressional attenuation is  0.1~dB$/\lambda$.

We solved each of the test cases with the parabolic equation solutions based on Eq.~(\ref{eq:modss}), Eq.~(\ref{eq:appec}), the combination of Eq. (\ref{eq:ss1}) and a layer with low shear speed that is used to model part of the water column, and the rotated coordinates approach.
In the low-speed layer, the compressional speed and density are the same as for water, there is no compressional attenuation, the shear speed is 10~m/s, and the shear attenuation is  10~dB/$\lambda$. 
There is no low-speed layer for $r<1.5$~km; the low-speed layer lies between $z=50$~m and the top of the sediment for $r>1.5$~km.
The low-contrast vertical interface at $r=1.5$~km is handled by conserving $u$ and  $\sigmaxz$.

\begin{figure}
\centering
\includegraphics[width=.6\textwidth]{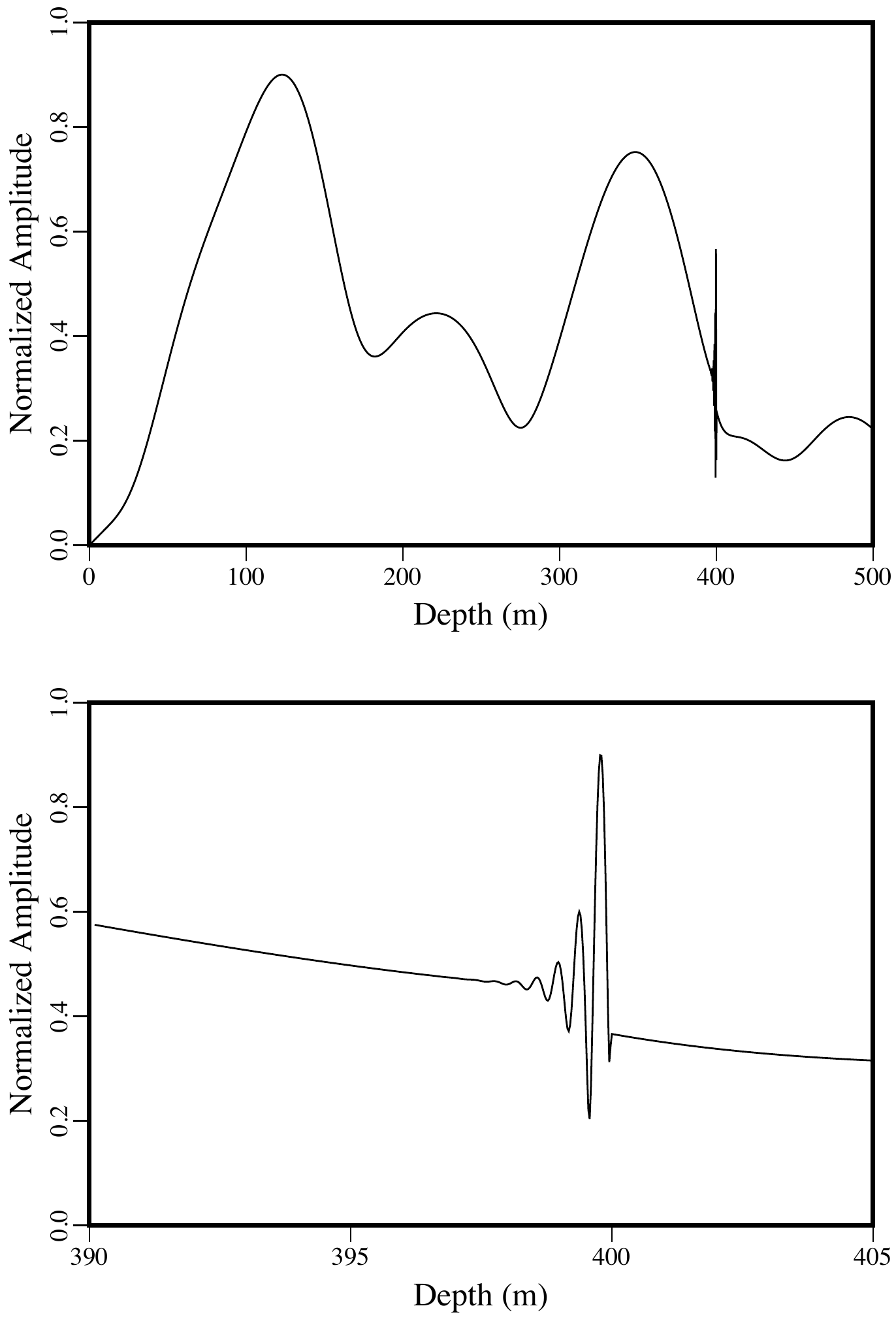}
\caption{Wide (top) and zoomed (bottom) views of the normalized amplitude of the horizontal displacement at $r=3$~km for the medium-contrast sediment. For $r>1.5$~km, the lower 350~m of the 400~m deep water column is modeled as a solid with very low shear speed. There are rapid variations in the horizontal displacement in a thin boundary layer above the interface. }
\label{fig:layerdisp}
\end{figure}

\begin{figure}
\centering
\includegraphics[width=.6\textwidth]{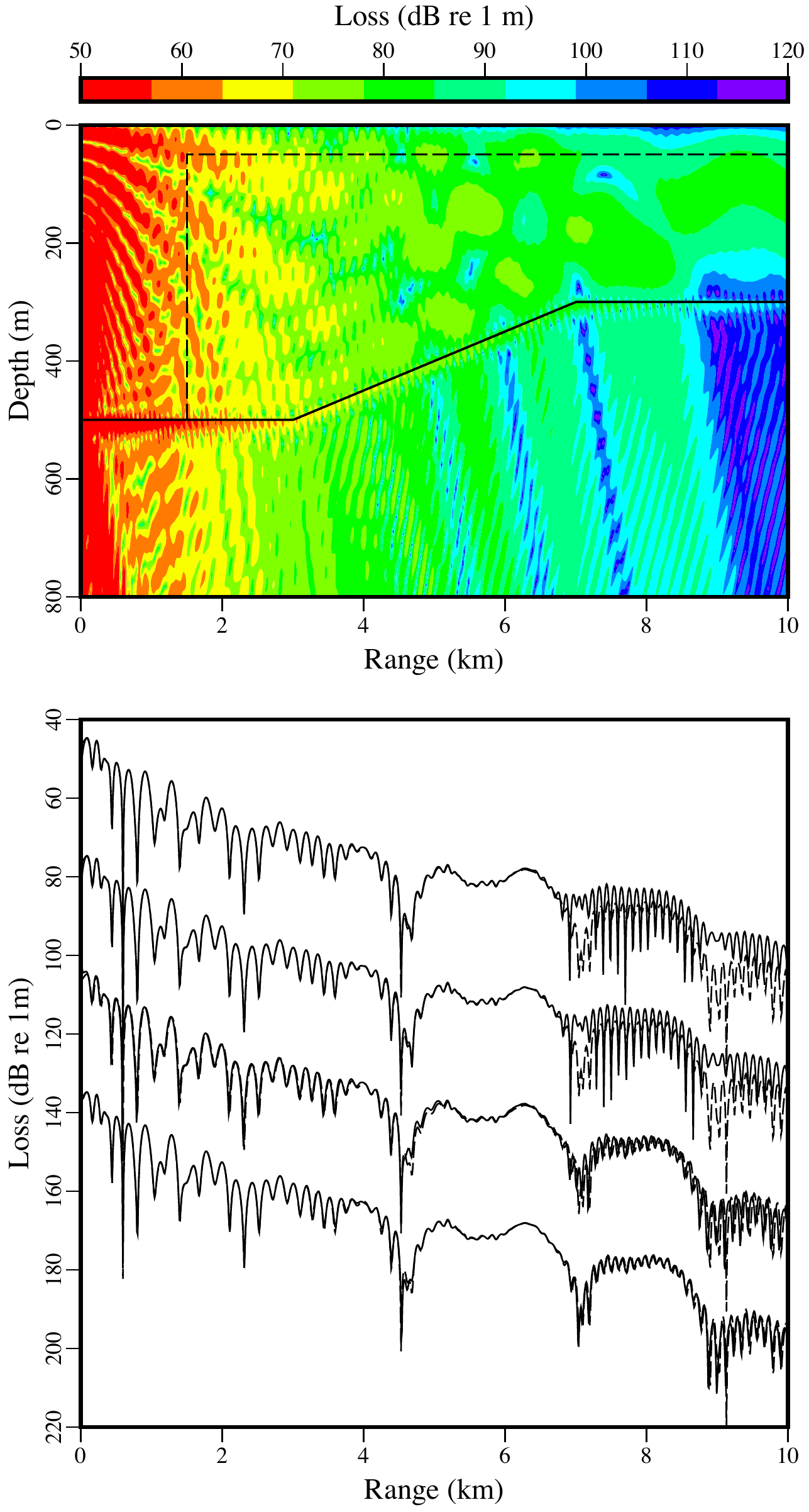}
\caption{Transmission loss color plot and curves at  $z=290$~m for a test problem with a Scholte wave that propagates up the slope. The dashed lines in the color plot mark the boundaries of the low-speed solid layer that is used to model a fluid. From top to bottom, the solid transmission loss curves are the solutions based on Eq.~(\ref{eq:modss}), Eq.~(\ref{eq:appec}), the low-speed layer, and rotated coordinates. The dashed curves correspond to a reference solution that was obtained with a finite-element model. }
\label{fig:scholteup}
\end{figure}

\begin{figure}
\centering
\includegraphics[width=.6\textwidth]{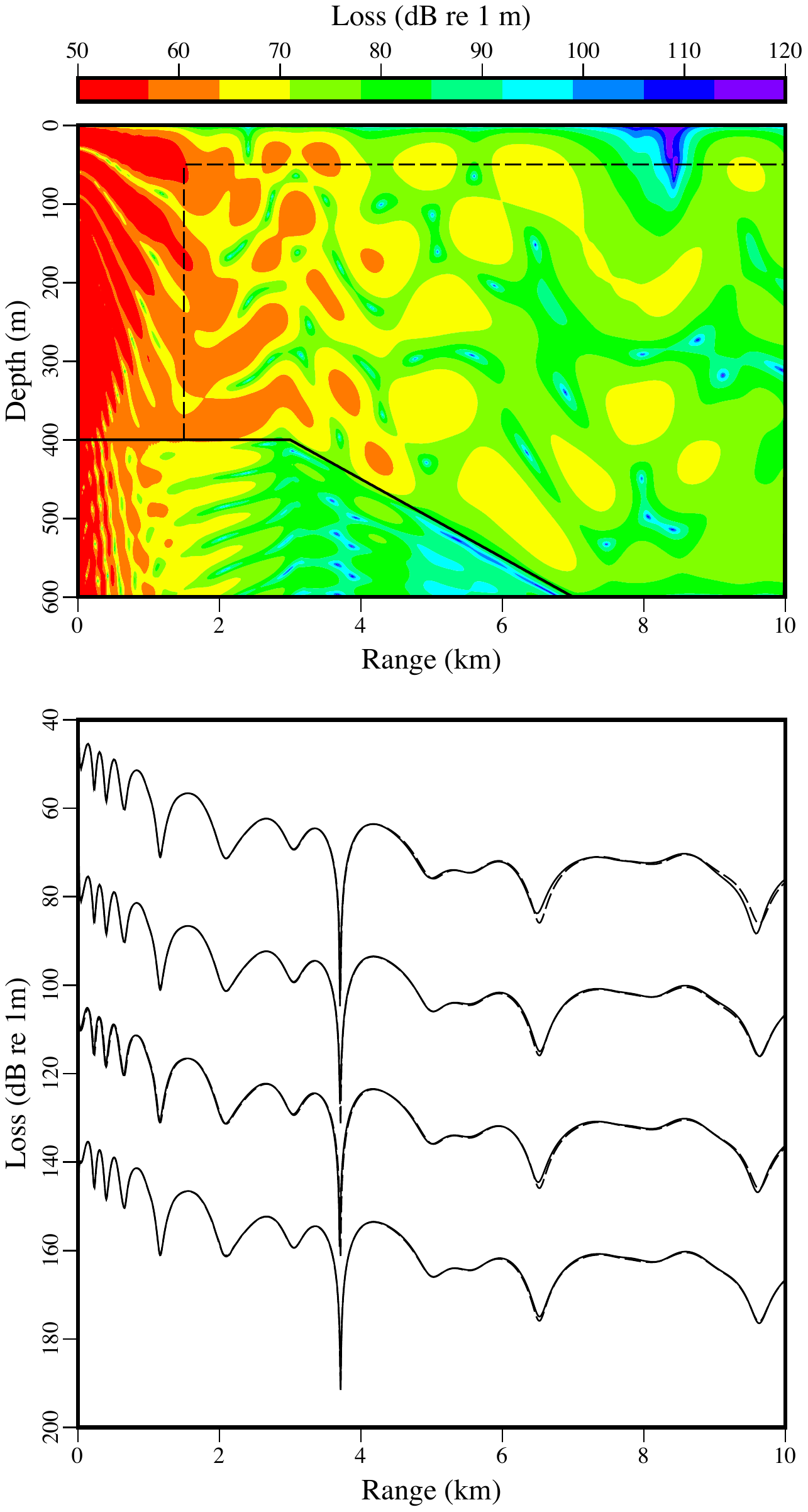}
\caption{Transmission loss color plot and curves at $z=180$~m  for the low-contrast downslope problem. The dashed lines in the color plot mark the boundaries of the low-speed solid layer that is used to model a fluid. From top to bottom, the solid transmission loss curves are the solutions based on Eq.~(\ref{eq:modss}), Eq.~(\ref{eq:appec}), the low-speed layer, and rotated coordinates. The dashed curves correspond to a reference solution that was obtained with a finite-element model.}
\label{fig:lcdown}
\end{figure}

\begin{figure}
\centering
\includegraphics[width=.6\textwidth]{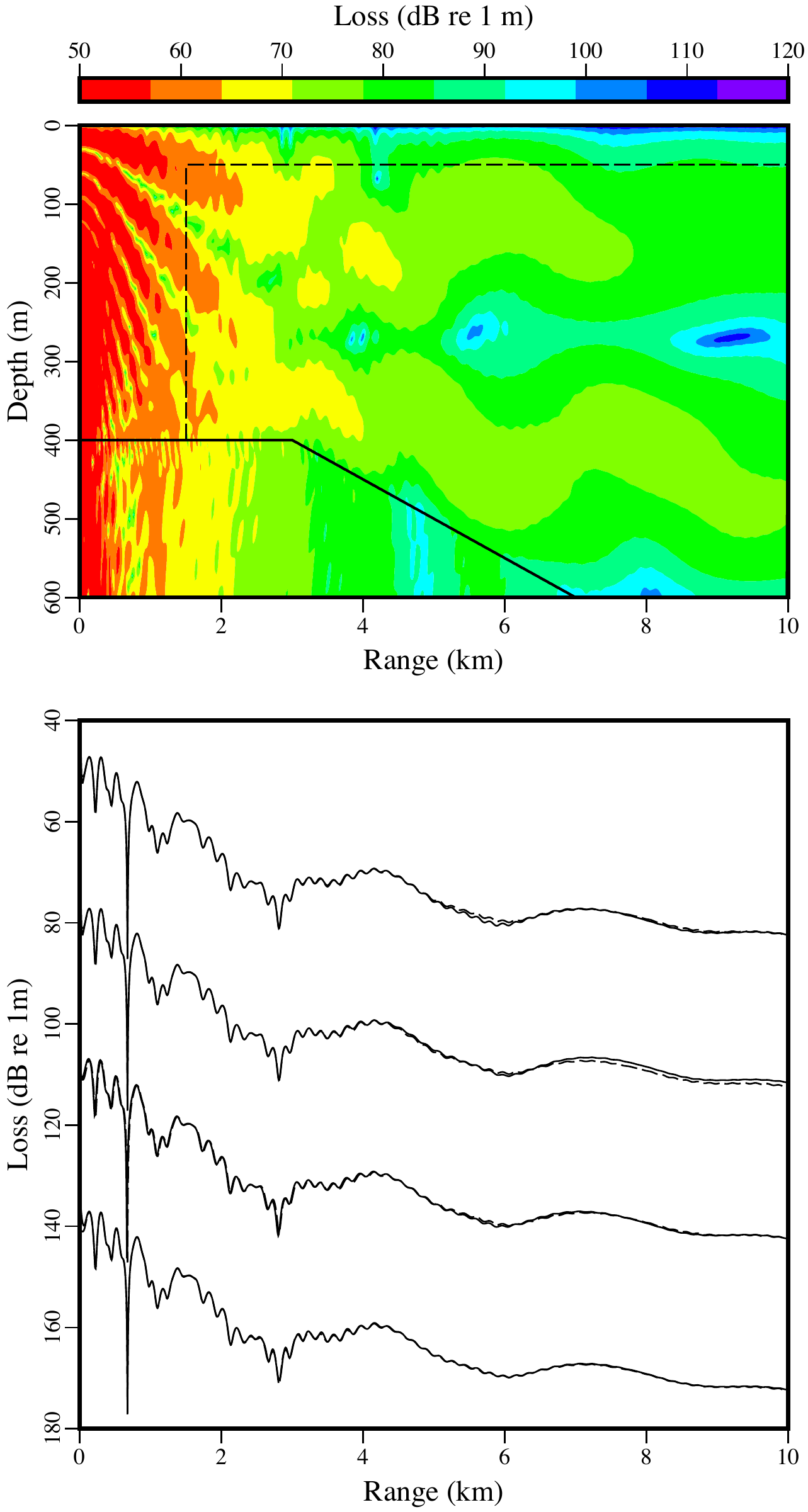}
\caption{Transmission loss color plot and curves at $z=180$~m  for the medium-contrast downslope problem. The dashed lines in the color plot mark the boundaries of the low-speed solid layer that is used to model a fluid. From top to bottom, the solid transmission loss curves are the solutions based on Eq.~(\ref{eq:modss}), Eq.~(\ref{eq:appec}), the low-speed layer, and rotated coordinates. The dashed curves correspond to a reference solution that was obtained with a finite-element model.}
\label{fig:mcdown}
\end{figure}

\begin{figure}
\centering
\includegraphics[width=.6\textwidth]{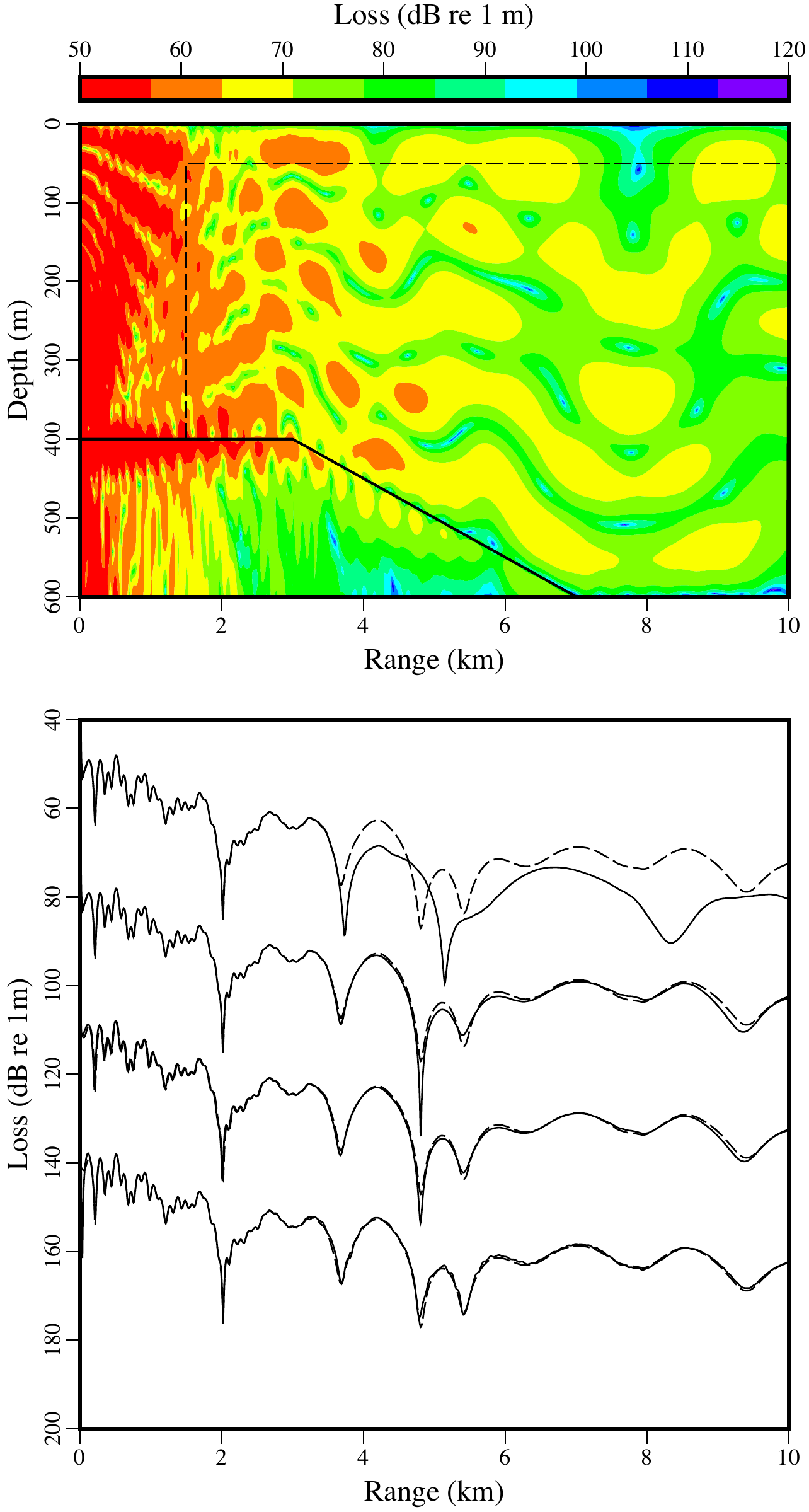}
\caption{Transmission loss color plot and curves at $z=180$~m  for the high-contrast downslope problem. The dashed lines in the color plot mark the boundaries of the low-speed solid layer that is used to model a fluid. From top to bottom, the solid transmission loss curves are the solutions based on Eq.~(\ref{eq:modss}), Eq.~(\ref{eq:appec}), the low-speed layer, and rotated coordinates. The dashed curves correspond to a reference solution that was obtained with a finite-element model.}
\label{fig:hcdown}
\end{figure}

\begin{figure}
\centering
\includegraphics[width=.6\textwidth]{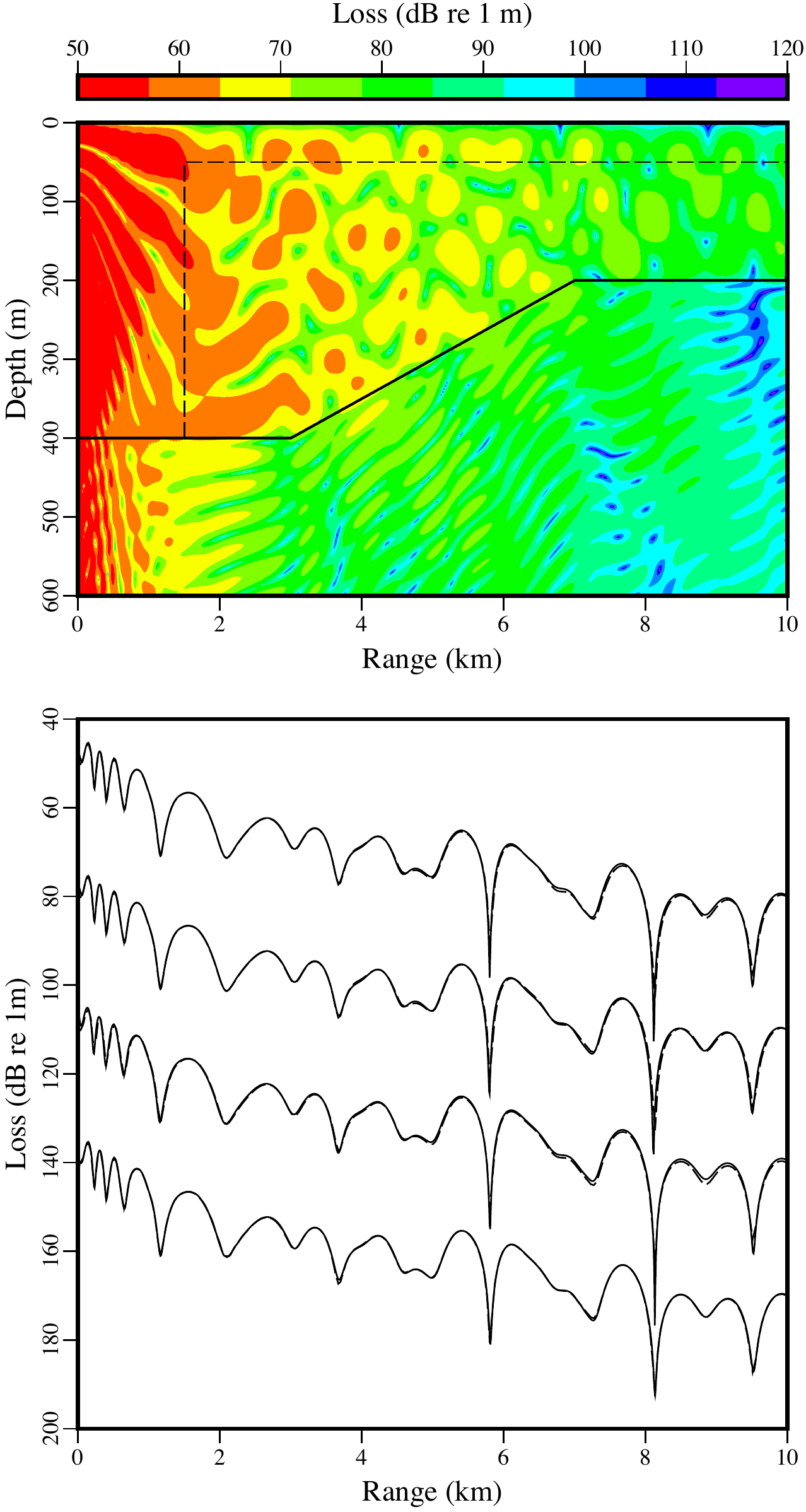}
\caption{Transmission loss color plot and curves at $z=180$~m  for the low-contrast upslope problem. The dashed lines in the color plot mark the boundaries of the low-speed solid layer that is used to model a fluid. From top to bottom, the solid transmission loss curves are the solutions based on Eq.~(\ref{eq:modss}), Eq.~(\ref{eq:appec}), the low-speed layer, and rotated coordinates. The dashed curves correspond to a reference solution that was obtained with a finite-element model.}
\label{fig:lcup}
\end{figure}

\begin{figure}
\centering
\includegraphics[width=.6\textwidth]{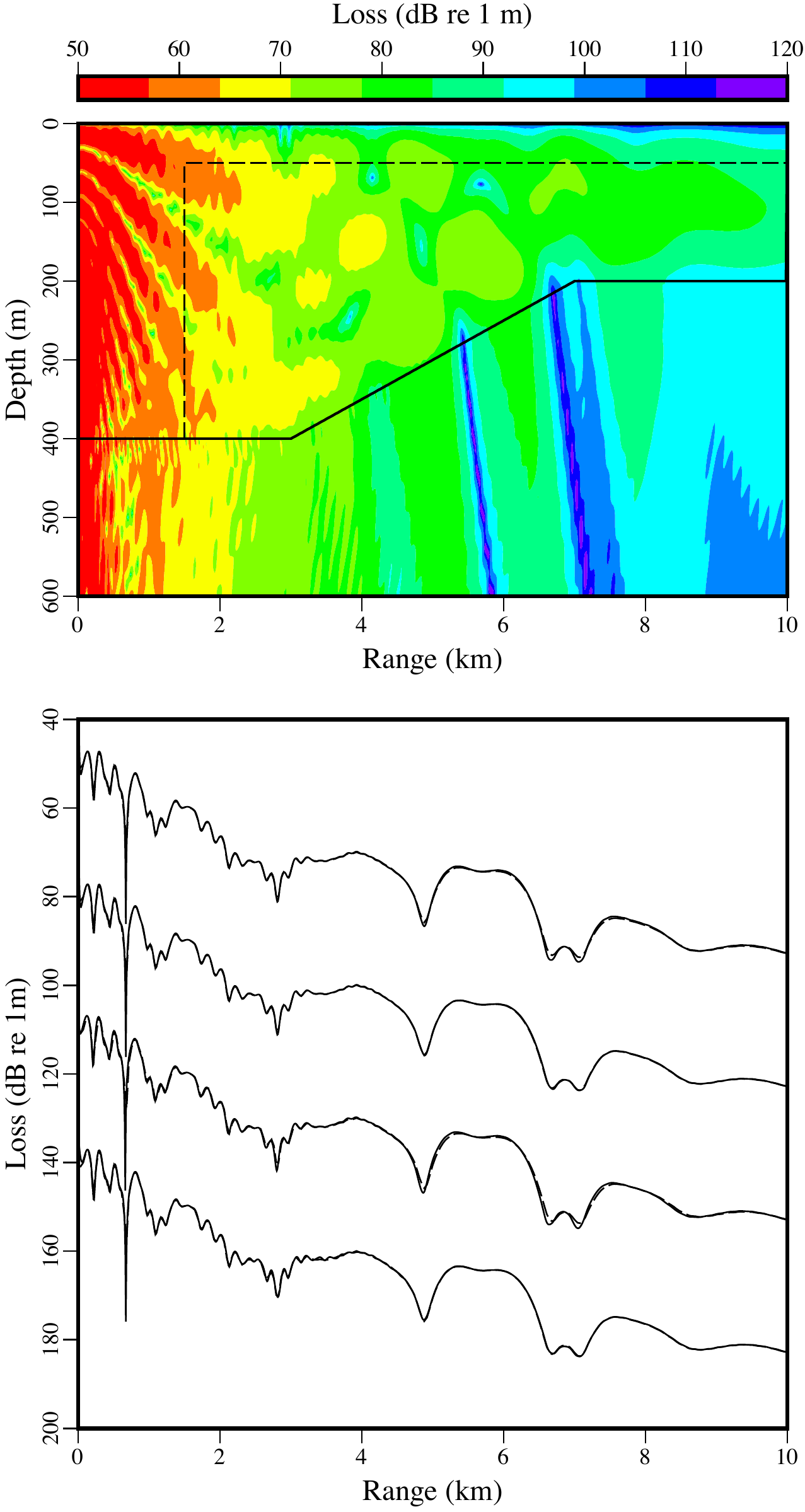}
\caption{Transmission loss color plot and curves at $z=180$~m  for the medium-contrast upslope problem. The dashed lines in the color plot mark the boundaries of the low-speed solid layer that is used to model a fluid. From top to bottom, the solid transmission loss curves are the solutions based on Eq.~(\ref{eq:modss}), Eq.~(\ref{eq:appec}), the low-speed layer, and rotated coordinates. The dashed curves correspond to a reference solution that was obtained with a finite-element model.}
\label{fig:mcup}
\end{figure}

\begin{figure}
\centering
\includegraphics[width=.6\textwidth]{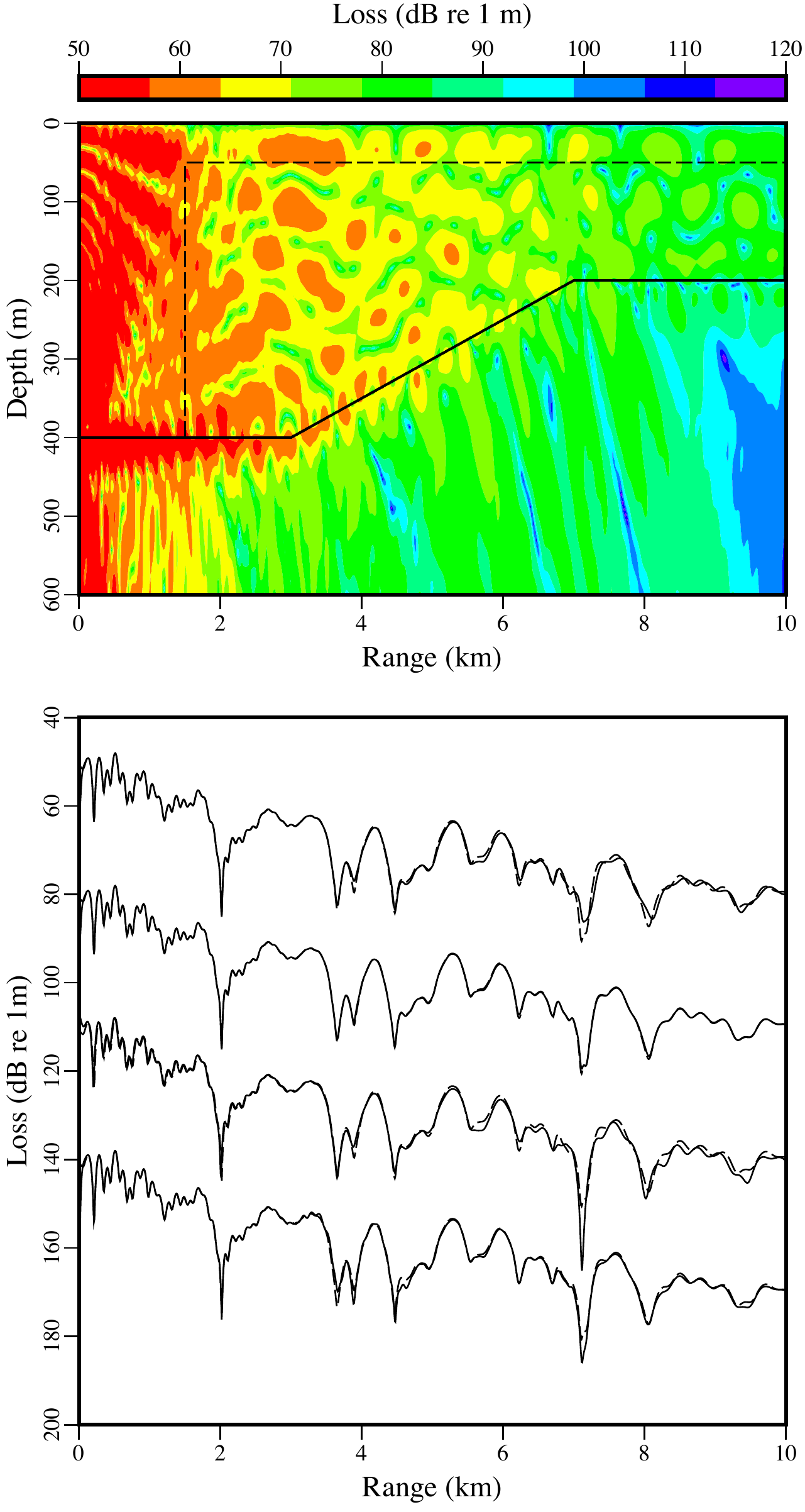}
\caption{Transmission loss color plot and curves at $z=180$~m  for the high-contrast upslope problem. The dashed lines in the color plot mark the boundaries of the low-speed solid layer that is used to model a fluid. From top to bottom, the solid transmission loss curves are the solutions based on Eq.~(\ref{eq:modss}), Eq.~(\ref{eq:appec}), the low-speed layer, and rotated coordinates. The dashed curves correspond to a reference solution that was obtained with a finite-element model.}
\label{fig:hcup}
\end{figure}

Appearing in Fig.~\ref{fig:layerdisp} are plots of $u$ at $r=3$~km  for the medium-contrast case. 
There are rapid variations in a boundary layer just above the interface at $z=400$~m; a slowly varying extrapolation of the slowly varying curve outside the boundary layer would not match the value of the curve in the sediment at the interface; this is consistent with the fact that u is not continuous across a fluid-solid interface. 
All four of the parabolic equation solutions appear in Figs.~\ref{fig:scholteup}--\ref{fig:hcup} as solid curves in the same order (shifted downward in increments of 30 dB from top to bottom) as they are listed above; the finite-element solutions appear as dashed curves; the boundary of the region with low shear speed is marked with a dashed curve in the color plots.

As shown in Fig.~\ref{fig:scholteup}, the solutions based on Eqs.~(\ref{eq:modss}) and (\ref{eq:appec}) break down for the interface wave case, but the other solutions are accurate for that case. 
As shown in Fig.~\ref{fig:hcdown}, the solution based on Eq.~(\ref{eq:modss}) breaks down for some downslope problems with a high-contrast sediment (it can also break down for some upslope problems). 
Although the solution based on Eq.~(\ref{eq:appec}) does not properly handle the propagation of a Scholte wave along a sloping interface, it otherwise appears to be accurate and stable for a large class of problems (in Ref.~\citen{collins2019b}), it was found to produce accurate solutions for problems involving ice cover. 
The solution that is based on a low shear speed layer provides accurate solutions for most of the cases, but there are some errors for the high-contrast cases. 
The solution based on rotated coordinates provides excellent results, but there are small errors for the high-contrast upslope case. 

\begin{figure}
\centering
\includegraphics[width=.6\textwidth]{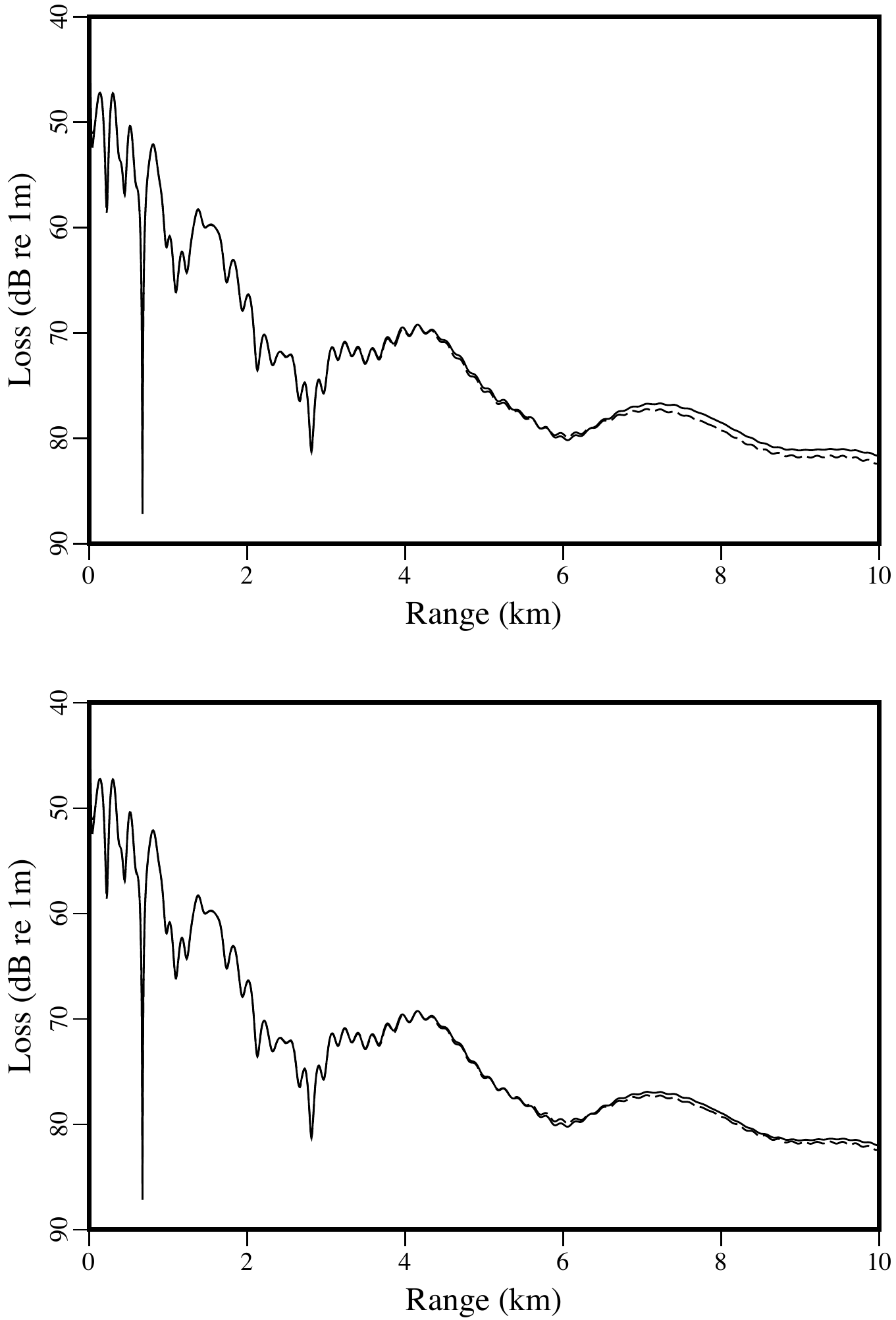}
\caption{Transmission loss color plot and curves at $z=180$~m  for the medium-contrast downslope problem. The solid lines correspond to parabolic equation solutions that were obtained using a film of low shear speed solid material on the rises with three (top) and four (bottom) grid points on the rises. The dashed curves correspond to a reference solution that was obtained with a finite-element model.}
\label{fig:mcdownfilm}
\end{figure}

\begin{figure}
\centering
\includegraphics[width=.6\textwidth]{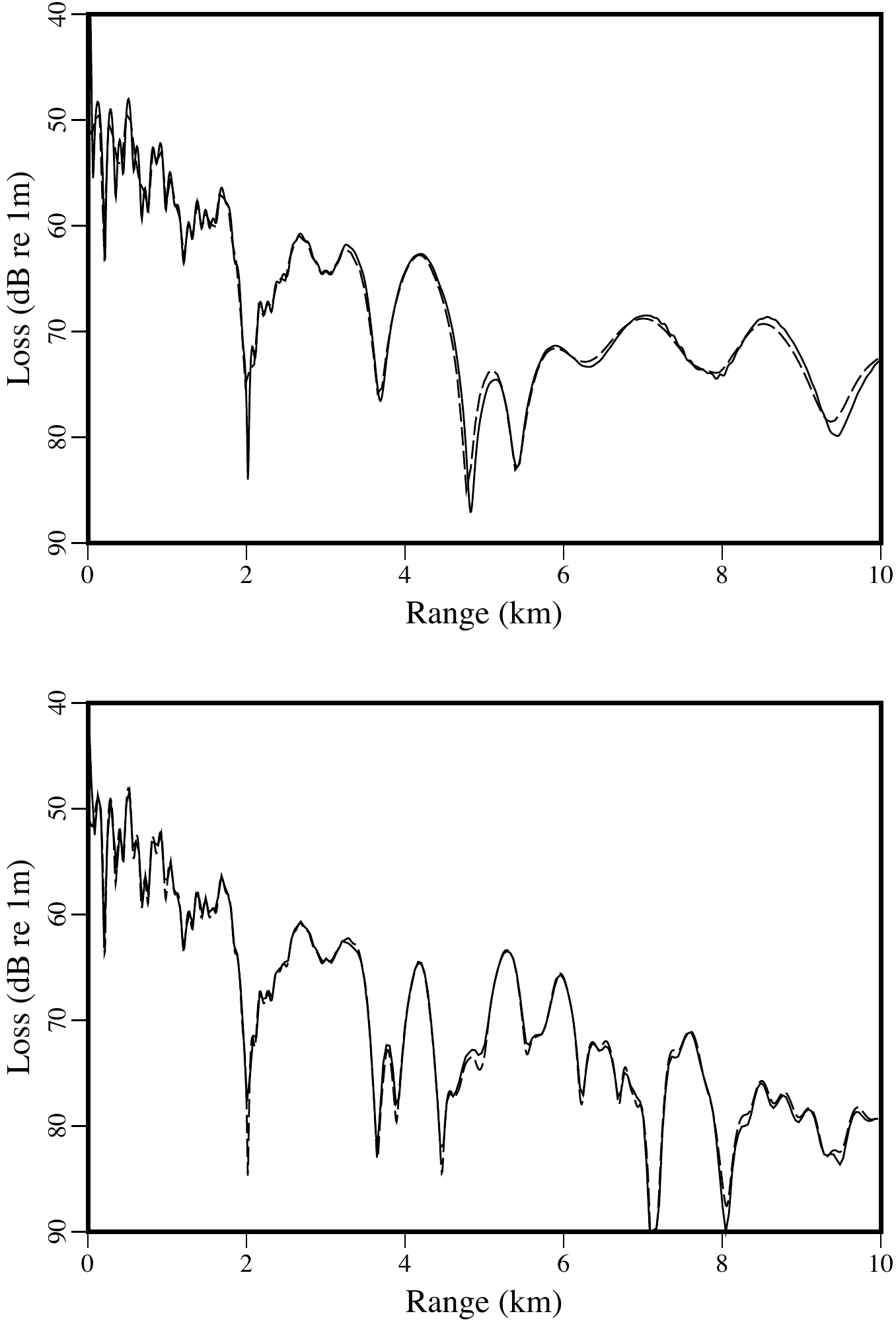}
\caption{Transmission loss color plot and curves at $z=180$~m  for the high-contrast downslope (top) and upslope (bottom) problems. The solid lines correspond to parabolic equation solutions that were obtained using a film of low shear speed solid material on the rises with three (top) and four (bottom) grid points on the rises. The solid lines correspond to parabolic equation solutions that were obtained with the corrected mapping approach. The dashed curves correspond to reference solutions that were obtained with a finite-element model.}
\label{fig:hccorrmap}
\end{figure}

The solutions for the medium-contrast downslope problem appearing in Fig.~\ref{fig:mcdownfilm} were obtained with the approach involving a film of low shear speed material on the rises. 
In the film, the shear speed is 500~m/s and the compressional speed and density are the same as in the water. 
The vertical interface on the incident side of the film is handled with Eq.~(\ref{eq:ss1}). 
The relatively low-contrast vertical interface on the transmitted side of the film is handled by conserving $u$ and $\sigmaxz$. 
The solution that was obtained using three grid points on the rises is not as accurate as the solution that was obtained using four grid points on the rises, which better approximate the shear speed profile in the film. 
Appearing in Fig.~\ref{fig:hccorrmap} are mapping solutions (with the phase correction described in Ref.~\citen{collins2000}), which are fairly accurate for the high-contrast cases. 

\section{Discussion}

	Several approaches for handling sloping fluid-solid interfaces have been developed and tested. The approaches based on Eqs.~(\ref{eq:modss}) and (\ref{eq:appec}) are not effective for problems involving Scholte waves. The approached based on Eq.~(\ref{eq:modss}) may break down when the contrast across the interface is sufficiently large. The approximate energy-conservation approach based on Eq.~(\ref{eq:appec}) performs well for a wide range of problems when there is no Scholte wave. The rotated coordinates approach properly handles Scholte waves and provides accurate solutions for a wide range of problems, but this approach would not be useful for problems involving frequent or continuous variations in slope. The approach based on Eq. (\ref{eq:ss1}) and a low shear speed layer above the interface is effective for a wide range of problems, including those involving Scholte waves, but this approach needs further development. The approach based on placing a thin film on the rises of stair steps is effective for handling some downslope problems. The corrected mapping solution is fairly accurate for many problems, and it is applicable to problems involving continuous variations in slope. When Scholte waves are not important, the approach based on Eq.~(\ref{eq:appec}) is perhaps the most attractive existing parabolic equation approach for handling range-dependent problems in seismo-acoustics. There is a need for further development and testing. It is possible that the most effective strategy will be to employ multiple approaches that have advantages for different types of problems. It is also possible that a different approach that renders existing approaches obsolete will be discovered. 

\section*{Acknowledgments}
This work was supported by the Office of Naval Research. AR is supported through NRL's Jerome and Isabella Karle Fellowship Program.

\end{document}